# Metamagnetism and the Fifth Order Susceptibility in UPt$_3$

B.S.Shivaram[1], Brian Dorsey[1], D.G. Hinks[2] and Pradeep Kumar[3]

[1]Department of Physics, University of Virginia, Charlottesville, VA. 22901.

[2]Material Science and Technology Division, Argonne National Labs, Argonne, IL. 60637.

[3]Department of Physics, P.O. Box 118440, University of Florida, Gainesville, FL.32611.

**ABSTRACT**

An enhanced susceptibility is a natural consequence of the "heavy fermion" (HF) state rendering the possibility of measurably large nonlinear susceptibilities. In recent work a universal behavior of the peaks observed in the linear ($\chi_1$) and the third order ($\chi_3$) susceptibility in HF metamagnets has been identified. This universality is well accounted for by a single energy scale model considering on-site correlations only. A prediction of this model is a peak in the fifth order susceptibility, $\chi_5$, as well. In the first measurements on a HF metamagnet, UPt$_3$ reported herein, we find that $\chi_5$ rather than attaining a peak, saturates at low temperatures and is positive. The thermodynamic implications of these towards the stability of the metamagnetic HF state are discussed.

The heavy electron metals exhibit a fascinating diversity of magnetic properties[1,2]. A significant fraction of them order antiferromagnetically, some even order as ferromagnets, and others remain nonmagnetic yet harboring within them other possible exotic types of magnetic order. Proposals of such exotic order include multipolar order such as quadrupolar (ferro and antiferro)[3], hexadecapolar[4], hidden order[5,6] etc. Higher order susceptibilities can be an indicator of multipolar order. At the same time, heavy electron materials also contain physics of Kondo compensation, and conceivably, nonlinear susceptibilities, caused by the response of Kondo-like compensation to a larger magnetic field. Irrespective of which of the scenarios above is valid the experimental measurement of successively higher order dc magnetic susceptibilities is useful.

The Wilson ratio, a measure of the strong correlations, is often close to unity in heavy electron materials. That ensures that the heavy effective mass also translates to a large susceptibility of the conduction electron system[7]. Beginning with such enhanced susceptibilities it may be surmised that the nonlinear components of the susceptibility might similarly be very large and easily accessed experimentally. This is indeed the case and a measurement of the third order susceptibility has been reported in a number of heavy electron systems[8,9,10,11]. We have recently reported the first measurements of $\chi_3$ in UPt$_3$ where we observe a peak in $\chi_3$ at a temperature approximately half the temperature where a peak in $\chi_1$ is observed[12]. We also noted that this scaling feature is observed in a number of other measurements reported in the literature. We have proposed a single energy scale model (see below) which accounts for the peak in $\chi_3$ as well as arriving at the correct relationship between the temperatures $T_3$ and $T_1$. Experimentally the ratio $T_3/T_1$ is close to 1/2 whereas the model gets the ratio to be 0.4. This model also predicts that there is a peak in the fifth order susceptibility, $\chi_5$ at a temperature lower than $T_3$.

In this letter we present the first measurements of the fifth order susceptibility in a heavy electron material. These measurements were performed at the National High Magnetic Field Laboratory, Florida on a single crystal of UPt$_3$ employing a capacitance torque magnetometer in fields up to 30 Tesla. Capacitance isotherms were measured as the magnetic field was swept for both orientations of the field with respect to the hexagonal crystallographic axes. The two capacitance isotherms were then deconvoluted to compute the magnetization separately for the field parallel to the c-axis as well as the a-axis of UPt$_3$. This deconvolution procedure has been described in detail in a recent application note[13]. The magnetization at low fields may be expanded in odd powers of the applied field H as:

$$M = \chi_1 H + \chi_3 H^3 + \chi_5 H^5 \qquad \text{Eqn.(1)}$$

Thus by plotting the ratio M/H as a function of $H^2$ we can discern both the third order and the fifth order susceptibilities from a quadratic fit to the experimental data with the intercept yielding the linear susceptibility $\chi_1$.

In figure 1 we show the magnetization along the a-axis (parallel to the basal plane) for nine different temperatures as identified in the individual panels. The magnetic isotherms at high temperatures start out similar to that of a conventional paramagnet with the moment tending to saturation gradually as the field is increased. This "normal" paramagnetic response means that M-$\chi_1$H < 0, making the third order susceptibility negative with a negative sign also for the fifth order susceptibility. As the temperature is lowered the isotherms "stiffen" implying $\chi_3$ is becoming less negative and with further decrease in temperature an upward curvature in the response develops thus signaling a positive trend in the nonlinear susceptibilities. At the lowest temperatures a sharp upward curvature in the magnetization at the metamagnetic field is apparent suggesting that either $\chi_3$ or $\chi_5$ or both are strongly positive. In order to separate out these two susceptibility components we follow the procedure explained above with plots of M/H vs. $H^2$. These plots are shown in the series of nine panels in figure 2. At high temperatures the curves in fig. 2 have a downward slope and curvature and yield negative values for $\chi_3$ and $\chi_5$. This negative slope quickly gives way to an almost flat response at T~14 K below which the opposite tendency starts to develop. At the lowest temperatures an upward curvature is clearly visible with the data for 2K in particular being a pure quadratic thus indicating that $\chi_3$ is nearly zero but $\chi_5$ is positive. The individual values of $\chi_5$ obtained through such plots are shown in fig.3. Here the data have been fit to $H^2$ values up to 250 $T^2$ which represents an order of magnitude increase in scale compared to that employed in our earlier work on $\chi_3$. The values of $\chi_5$ are negative at high temperatures, turn positive at lower temperatures crossing zero at approximately 10 K. $\chi_5$ continues to remain positive at the lowest temperatures in this study with no indication of a peak. For purposes of comparison and to establish a calibration scale we also show the intercept $\chi_1$ obtained through the fits in the upper right inset of fig.3. Also shown here, for comparison, are the more quantitative precise measurements derived from SQUID magnetometry. The torque magnetometry signals we measure as shown in fig. 2 are uncalibrated and thus we have used an appropriate scale factor to match the values of $\chi_1$ in fig. 3 with those from SQUID magnetization data[12].

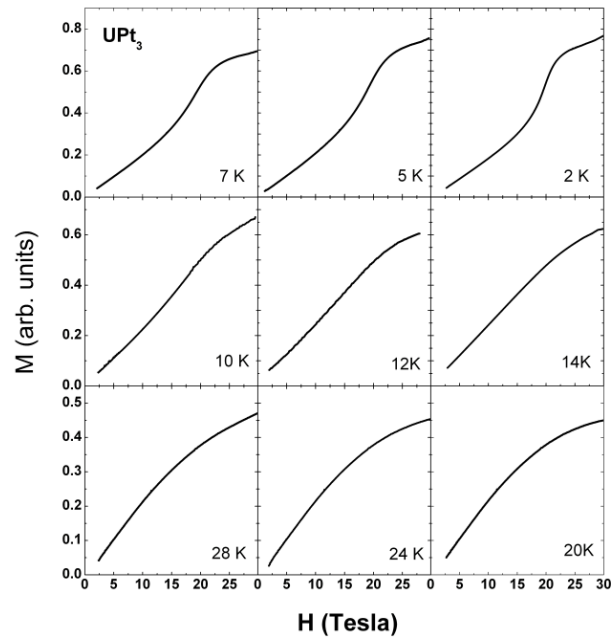

*Fig.1: Shows the magnetization isotherms of UPt$_3$ with field applied along the a-axis. A "normal" paramagnetic response observed at high temperatures with a negative bending gives way to an upward curvature at lower temperatures signaling the rise of higher order susceptibilities to positive values prior to the metamagnetic transition itself at 20 T.*

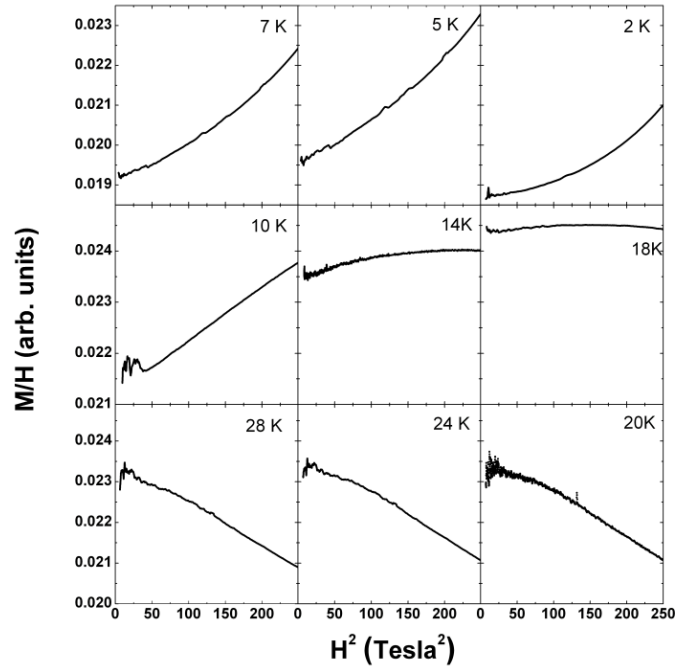

*Fig. 2: Shows the nonlinear part of the magnetic response after subtracting out the linear term. The behavior at high temperatures (lower panels) is almost linear and with sloping down thus implying a negative $\chi_3$. Around 14 K $\chi_3$ turns positive with a definite indication of a -ve $\chi_5$. The response at 10 K is largely a straight line indicating a large positive $\chi_3$ and $\chi_5$ about to cross over to the positive side. The response at lower temperatures clearly shows both $\chi_3$ and $\chi_5$ are positive with the exception of the data at 2 K which is a perfect parabola indicating a near zero $\chi_3$.*

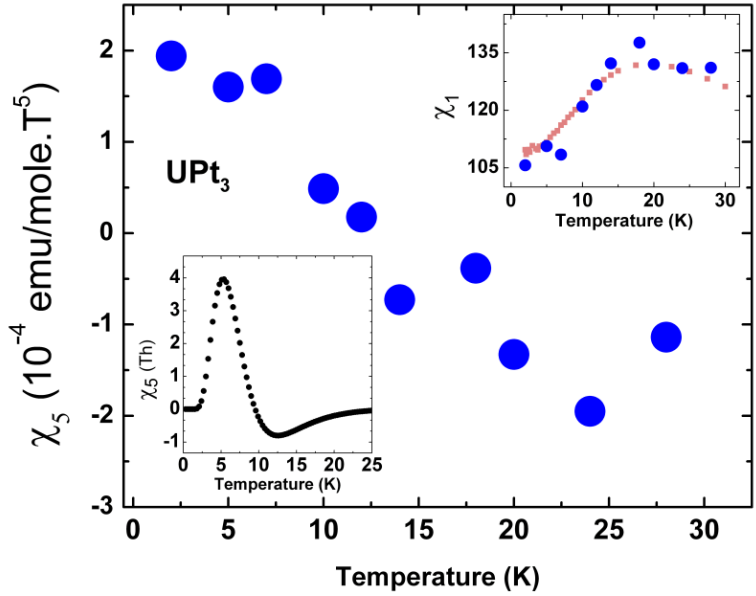

*Figure 3: Shows the fifth order susceptibility, $\chi_5$ in $UPt_3$ obtained from the quadratic term in fits to the data shown in fig.2. $\chi_5$ is negative at high temperatures and increases monotonically as $T \rightarrow 0$. The lower left inset shows the calculated $\chi_5$ as derived in the single energy scale model. The upper right inset shows the intercepts in fig.2 appropriately scaled (blue dots) to match the linear susceptibility obtained from SQUID measurements (red dots).*

In the work that was reported in ref.(12) the peak temperatures $T_3$ and $T_1$ followed the relation $T_3 \cong T_1/2$. Seeking a simple understanding of this relation (as well as the earlier correlations between the critical field and $T_1$ and the critical field and inverse of the peak linear susceptibility) we were led to a minimal one energy scale model which contains (a) metamagnetic transition and (b) an opportunity to calculate $\chi_3(T)$, which seemed to capture the essential details of the observed temperature dependence. The materials we considered are primarily Kondo lattices likely with incompletely compensated local moments which are also in a crystalline lattice and therefore with electronic energy levels subject to the crystalline electric fields. Our model at the current stage does not depend on the exact crystalline structure. We also cannot assert with any certainty that the metamagnetism is a form of shakeoff of the Kondo cloud. However it seems certain that we need only one energy scale and the minimal model produces the suggestive functional forms of all measured quantities.

In the minimal model we start with the Hamiltonian $H = \Delta S_z^2 - \gamma S_z B$, where $\Delta$ sets the energy scale and could represent the separation between the singlet ground state and the lowest excited state. In this model the quantum spins are treated as discrete. When the magnetic field is parallel to the quantization axis (z-axis) the susceptibilities are calculated as:

$$\chi_1 = \frac{\gamma^2}{\Delta} \frac{1}{\tau} \frac{1}{1+A} \tag{2}$$

$$\chi_3 = \frac{\gamma^4}{3!\Delta^3} \frac{1}{\tau^3} \frac{A-2}{(1+A)^2} \tag{3}$$

$$\chi_5 = \frac{\gamma^6}{5!\Delta^5} \frac{1}{\tau^5} \frac{A^2-13A+16}{(1+A)^3} \tag{4}$$

where A= 0.5 $e^{1/\tau}$ and $\tau$= $k_BT/\Delta$. The principal features of $\chi_5(T)$ are similar to those of $\chi_3(T)$. At high temperatures, $\chi_5(T) < 0$. It has a peak at $T_5 = 0.17\Delta$. The value of $\Delta$ for UPt$_3$ (obtained from the measured values of $H_m$, $T_1$ and $T_3$) is 30 K and hence we can expect a peak at 5 K[14]. Our data clearly shows a rise in $\chi_5$ but no fall off below that temperature. There is no indication of a peak. It is noteworthy that both $\chi_1$ and $\chi_5$ saturate and are positive at the lowest temperatures measured in UPt$_3$ while the intermediate susceptibility $\chi_3$ is close to zero.

For further analysis we begin with a general expansion of the magnetic free energy for B and M parallel to the z-axis:

$$F = - B.M + a_2 M^2 + a_4 M^4 + a_6 M^6 + .... \tag{5}$$

In such an expansion the coefficients $a_2$, $a_4$, $a_6$ can be related to the susceptibilities.

$$a_2 = \frac{1}{2\chi_1} \tag{6}$$

$$a_4 = -\chi_3/4\chi_1^4 \tag{7}$$

$$a_6 = \frac{1}{\chi_1^7}[3\chi_3^2 - \chi_5\chi_1] \tag{8}$$

When all coefficients are positive, the minimum of this free energy is M=0. Since $\chi_1 \geq 0$ it follows that $a_2 > 0$ and there is no transition here driven by the quadratic term. From our measurements though we see that $\chi_3(T)$ changes sign and starting from a negative value at high temperatures ($a_4 \geq 0$) it becomes positive indicating a potentially discontinuous transition in M. The next coefficient $a_6$, if positive would keep the free energy bounded. In terms of nonlinear susceptibilities, $a_6$ is expressed in Eq.(8) and is seen to be negative for T < 12 K. It is clear from the data that the positive and growing value of $\chi_5$ below about 10 K has a critical effect on the condition expressed by eqn.(8). The fact that $a_6$ is about to vanish should have other consequences and the situation is reminiscent of the B-A transition in superfluid He-3[15]. The breakdown of the condition as shown in fig.5 implies one or both of the following two scenarios can occur: (a) higher order terms such as $a_8$ are required for keeping the free energy bounded (b) domains are formed below the instability temperature thus necessitating gradient terms in the free energy[16]. We note that there is a precedence for the presence of antiferromagnetic domains in UPt$_3$ in particular[17,18,19] and in strongly correlated metamagnets in general[20]. Also as noted above experimentally the T→0 value of $\chi_1$ is greater than zero and this is at variance with the single energy scale model. It is quite likely that resolving this issue would also result in additional terms in the free energy thus preserving a bounded free energy.

In conclusion, we have measured for the first time the fifth order nonlinear magnetic susceptibility $\chi_5(T)$ in the heavy fermion compound UPt$_3$. This is shown in fig.3 along with the expectation from a single energy scale model which we had developed earlier (shown in the inset). Whereas the model predicts a vanishing $\chi_5(T=0)$ the experiments show saturation and no turnover into a peak. The experimental data taken together with a polynomial expansion of the magnetic free energy in powers of the magnetization M confined to 6$^{th}$ order imply a possible thermodynamic instability around T=10K. Detailed measurements of higher order susceptibilities at low temperatures in other heavy fermion systems would be valuable in establishing whether the implied instability is a generic feature in itinerant metamagnets.

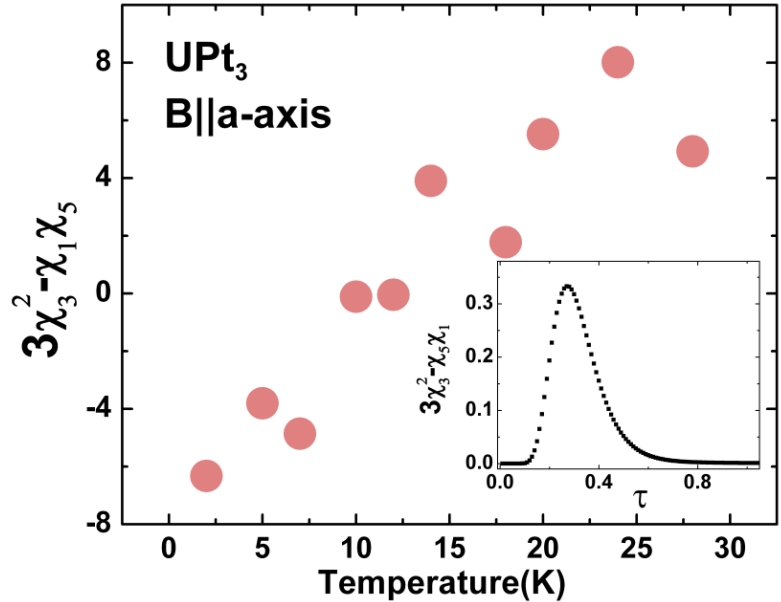

*Fig.4: The quantity on the vertical axis is a measure of the stability of the metamagnetic state. At low temperatures (~10K) this quantity assumes a negative value thus implying that additional terms in the magnetic free energy expansion need to be considered in order to ensure thermodynamic stability. The inset shows the criterion eqn.(8) as computed from the minimal model. The computed response never crosses zero.*

**Acknowledgements:** The work at the University of Virginia was supported through grant NSF DMR 0073456. Work at the NHMFL was supported by the National Science Foundation and the State of Florida. We acknowledge the invaluable assistance of Eric Palm, Tim Murphy and Donovan Hall at the NHMFL, Tallahassee, Florida. We acknowledge useful conversations with John Ketterson, Brian Maple and Jim Sauls.